\def\beq{\begin{equation}}
\def\eeq{\end{equation}}
\def\beqa{\begin{eqnarray}}
\def\eeqa{\end{eqnarray}}
\def\eg{${\it e.g.}$}
\def\ie{${\rm i.e.}$}
\def\subbeqa{\begin{subeqnarray}}
\def\subeeqa{\end{subeqnarray}}

\documentstyle[seceq,preprint]{jpsj}
\def\runtitle{Charge Excitations in Doped Mott Insulator 
in One Dimension}
\def\runauthor{Michiyasu {\sc Mori} and Hidetoshi {\sc Fukuyama}}

\title
{
Charge Excitations in Doped Mott Insulator in One Dimension
}

\author
{ 
Michiyasu {\sc Mori}
and Hidetoshi {\sc Fukuyama}
}

\inst
{
Department of Physics, the University of Tokyo, 7-3-1 Hongo, 
Bunkyo-ku, Tokyo 113\\
}

\recdate
{
\today
}

\abst
{The doped Mott insulator in one dimension has been studied 
 based on the phase Hamiltonian with the Umklapp scattering process,  
in which the charge degree of freedom is described by 
the quantum sine-Gordon model. 
The well-known equivalence between the quantum sine-Gordon model and 
the massive Thirring model for the spinless fermion makes it clear that the 
Mott-Hubbard gap originates from the Umklapp scattering process as was 
indicated by Emery and Giamarchi. 
Compressibility,  density-density correlation function,  
frequency dependence of optical conductivity and Drude weight 
have been calculated in the presence of the impurity scattering treated  
in the self-consistent Born approximation. It is seen that there exists a crossover behavior 
in the spectral weight of charge excitations: 
the acoustic mode is dominant in small wave number region while the optical excitations 
across the Mott-Hubbard gap lie in large wave number region and that 
this crossover wave number is reduced as the Mott transition is approached.}

\kword
{
Mott Transition, Umklapp scattering process, quantum sine-Gordon model
Massive Thirring model, compressibility, Drude weight, optical conductivity,
impurity scattering
}

\begin{document}
\sloppy
\maketitle

\section{Introduction}
 The Mott transition had so far been generally studied as metal-insulator transition which 
is caused by changing the strength of interaction (or the band width)
\cite{hubbard,br} at a particular carrier density of half-filling.  Since the 
discovery of high-${\it T_{\rm c}}$ material, however, the behavior as the band filling (or doping rate) is varied in the 
vicinity of Mott insulator is the central issue, \ie, doping into the Mott insulator 
(doped Mott insulator).
The high-${\it T_{\rm c}}$ superconductivities can be considered as one of ground states
of this doped Mott insulator.
In the metallic state of this doped Mott insulator, anomalous behaviors are observed
in transport and magnetic properties and then this state is called 
anomalous metallic state.
The origin of those anomalies are considered to arise from
characteristic nature of electronic states near the Mott transition. 

  For theoretical studies on this anomalous metallic phase, 
the Hubbard model has been frequently used. 
Although this model is considered to be the 
simplest Hamiltonian  containing the essence of  Mott transition, 
little is known about 
electronic properties near the transition as the doping rate is varied.
In one dimension, 
the Hubbard model has been solved exactly by use of Bethe Ansatz\cite{liebwu} 
and anomalous behaviors of various 
quantities,  compressibility, Drude weight etc., have been disclosed.
\cite{usuki,shastry}
Further development has been achieved recently by the conformal 
field theory and  asymptotic form of 
correlation functions have been determined.\cite{kawakami,korepin,schulz} 
These correlation functions behave as the power law which is characteristic to
the Tomonaga-Luttinger liquid\cite{tomonaga,luttinger} 
 where there exist excitations of acoustic modes 
both for charge and spin degrees of freedom in the limit of long wavelength.
However, the detailed nature of the excitation spectrum has not been 
explored with this framework. 

On the other hand, these properties can be explored by the 
bosonization.\cite{emery,solyom,fukuyama}
The system far away from half-filling is in the Tomonaga-Luttinger 
regime.\cite{tomonaga,luttinger}  As for a Mott insulator, \ie, at half-filling, Emery has shown  
that the Mott-Hubbard gap is seen to result from the Umklapp scattering 
process which is neglected in the Tomonaga-Luttinger 
regime.\cite{emery1,emery2}
The Mott transition caused by carrier doping, \ie, the doped Mott insulator, has been studied by 
Giamarchi.\cite{giamarchi} In this study, optical conductivity was calculated within 
the second order of Umklapp scattering process by use 
of memory-function approximation and renormalization group 
method.\cite{giamarchi,giamarchi2,giamarchi3} The Drude weight and 
its critical behavior toward the Mott transition has been investigated    
 by Giamarchi\cite{giamarchi} and Emery.\cite{emery3} 
These studies on the optical conductivity are concerned with charge excitation 
at $q=0$. On the other hand, 
charge excitation for finite q and 
$\omega$ near the Mott transition has been  disclosed in 
our previous work\cite{mori} where we have found interesting features of spectral weight 
of charge excitations; \ie, the crossover behavior where the 
acoustic mode, which is the property of Tomonaga-Luttinger liquid, has a large 
spectral weight in the region of the small wave numbers while the optical mode, 
which originates 
from  the inter-band excitation across the Mott-Hubbard gap, is weighted in 
the region of large wave numbers. The characteristic wave number of this crossover 
is reduced to zero as the metallic state merges to the insulating state. 

In this present work, we will study in detail effects of disorder on charge excitations 
in the doped Mott insulator. 
To investigate such effects, 
the phase Hamiltonian\cite{emery,solyom,fukuyama} has been employed, where 
the charge degree of freedom  is described by
the quantum sine-Gordon model  (QSG model) with misfit parameter 
representing the proximity to the half-filling.\cite{fukuyama,suzumura}
The phase Hamiltonian has been originally derived by the bosonization 
to describe the low energy excitations for correlated systems 
 in the limit of weak coupling \cite{suzumura}  and has been considered to be 
qualitatively applicable to the regime of strong coupling by choosing properly
 the constants of the Hamiltonian.\cite{haldane,penc}
In the classical treatment of this model, it is known that discommensuration lattice
(soliton lattice), which will be identified with 
particles corresponding to the charge degree of freedom,\cite{emery2} is created as the misfit 
parameter is varied and that there exists a transition between the commensurate 
and incommensurate phase as clarified by McMillan.\cite{mcmillan} 

We will study quantum mechanically the spectra of charge excitation 
by mapping QSG model onto the massive Thirring model (MT model).
\cite{coleman,mandelstam,okwamoto,tm,hida}
This useful mapping was originally applied by Luther and Emery
to the study of spin dynamics in 1D electron systems with backward scattering 
process\cite{luther,lee} and later to the Mott 
transition\cite{emery2,giamarchi,giamarchi2,giamarchi3,emery3,mori} as 
mentioned above. 
However, the detailed study of charge excitation including disorder 
has not been carried out especially for the case where the Mott transition 
is approached by the carrier doping.
By making use of the mapping and including disorder  in terms of 
spinless Fermion,   we will study the compressibility, density-density correlation 
function and  optical conductivity in the presence of impurity scattering.

The structure of this paper is as follows. To make this paper complete, 
we summarize the results of some previous works 
by Emery, Giamarchi and ourselves in \S 2 and \S 3. 
In \S 2, we introduce the
 massive Thirring model to describe the charge degree of freedom. This
model is derived from the phase Hamiltonian by using the bosonization.
In \S 3, the charge excitation spectrum and optical conductivity are calculated in a 
clean system. In \S 4, we introduce an impurity scattering to investigate these 
quantities in disordered systems. \S 5 is for conclusion and 
discussion. We confine ourselves to absolute zero temperature in this paper.
\section{Hamiltonian and One Particle Properties}
\subsection{Phase Hamiltonian and Classical Picture of Excitations}
  In one dimensional electron system,
the low energy excitation of charge and spin degrees of freedom are independent of
each another (spin-charge separation). Therefore, the phase Hamiltonian comprises
 two parts as follows;
\beq
H_\rho  = \int {\rm d}x \left[A_\rho (\nabla\theta(x))^2+
B_\rho \cos(2\theta(x)-q_0x)+C_\rho P^2\right], \\
\label{phasecharge}
\eeq
\beq
H_\sigma=\int {\rm d}x \left[A_\sigma(\nabla\phi(x))^2+
B_\sigma \cos(2\phi(x))+C_\sigma M^2\right],
\label{phasespin}
\eeq
where
\beq
[\theta(x),P(y)]={\rm i}\delta(x-y),\nonumber
\label{commu}
\eeq
\beq
[\phi(x),M(y)]={\rm i}\delta(x-y).
\eeq
 $\theta(x)$ and $\phi(x)$
describe the fluctuation of charge and spin degree of freedom, respectively.
The parameters
$A_\rho$, $B_\rho$, $C_\rho$, $A_\sigma$, $B_\sigma$ and $C_\sigma$ depend 
on the coupling constants and the filling of band. The charge excitations, which 
are our main interest in this paper, are described by $H_\rho$, eq.(\ref{phasecharge}), 
where the $B_{\rho}$-term is due to  Umklapp scattering with $q_0=G-4k_0$, 
the misfit parameter. Here, $G$ is the reciprocal lattice vector 
and $k_0=\mu/v_{F0}$, with $\mu$ and $v_{F0}$ being the chemical potential and
 the Fermi velocity 
of the noninteracting electrons, respectively. In the case of the noninteracting 
system, $k_0$ is the Fermi momentum. 
This misfit parameter controls the doping rate, \ie, the number of hole. 

In order to understand the physical implication of QSG model, 
eq.(\ref{phasecharge}), as  a function of the misfit parameter, 
we first analyze the model semi-classically by assuming  that the field operator, 
$\theta(x)$, 
consists of classical solution, $\theta_c(x)$, and fluctuation around this solution,  
$\hat{\theta}(x,t)$, as $\theta(x)=\theta_c(x)+\hat{\theta}(x,t)$.
Here $\theta_c(x)$ and  $\hat{\theta}(x,t)$ satisfy the following equations, 
respectively,
\beq
A_\rho (\nabla\theta_c(x))^2+B_\rho \cos(2\theta_c(x))=D,
\label{classical}
\eeq
\beq
\nabla^2\varphi(x)+\left[(m^2+\tilde{\omega}^2)-2m^2{\rm sn}^{2}(mx/k; k)
\right]\varphi(x)=0,
\label{fluctuation}
\eeq
where we assume $\hat{\theta}(x,t)=\varphi(x) e^{{\rm i}\omega t}$ and 
\beqa
m^{2}&=&\frac{2B_\rho}{A_\rho},\nonumber\\
k^{2}&=&\frac{2m^2}{m^2 + 2D/A_\rho},\nonumber\\   
\tilde{\omega}^2&=&\frac{\omega^2}{4A_{\rho}C_{\rho}}.
\eeqa
In eq.(\ref{fluctuation}),  
${\rm sn}^{2}(mx/k; k)$ is the Jacobian elliptic function with the parameter $k$.
An integral constant, $D$,  is to be evaluated by the solution, $\theta_c(x)$, which 
minimizes the energy per unit length, 
$\epsilon$, given by, 
\beq
\epsilon=\frac{1}{2\pi l}\int_{0}^{2\pi l}{\rm d}x 
A_\rho (\nabla\theta_c(x)+q_0)^2+
B_\rho \cos(2\theta_c(x)),\nonumber
\eeq
where $l$ is defined as, $2\pi l = (2k/m)K(k)$, 
and identified to a mean distance between discommensuration(DC)s. $K(k)$ is the 
first kind complete elliptic integral with parameter k. 
In the classical solution, $\theta_c(x)=\pi/2-{\rm sn}(mx/k; k)$, 
there exist a transition between the commensurate(C) and incommensurate(IC) phase 
as is shown in Fig.1 for $m=1$ and some choices of $k$.\cite{mcmillan}
Since the fluctuation of number density of charge, $n(x)$, in the 
long wave length limit ($q\ll2k_0$) is ,
\beq
n(x)\sim\frac{\nabla\theta(x)}{\pi},
\label{soliton}
\eeq
the DCs of $\theta_c(x)$ are identified  as particles.\cite{emery2} Thus  
the commensurate (incommensurate) phase can be considered to be 
an undoped (doped) state. In the IC phase 
near the Mott transition,  the DC lattice is created and 
its excitation spectrum, which is the eigenvalue of eq.(\ref{fluctuation}) and
 is shown in Fig.2 for $m=1$ and $k^2=0.5$,\cite{mcmillan,sutherland} 
 is considered to be vibrations of DC lattice; 
the acoustic mode of the DC lattice in the first Brillouin zone and the 
optical mode in the extended zone.\cite{mcmillan,sutherland}
In this classical treatment, there exists a critical wave number, $\pi/l$, which 
separates the region of the acoustic and optical mode.
 This $\pi/l$ characterizes how  the metallic 
state merges into the Mott insulator. At the C-IC transition, the acoustic
mode disappears and only the optical mode survives.  These characteristic features 
are held even in fully quantum mechanical treatment, as we will see. 
Although general features of excitation spectrum can be understood by the 
classical picture, the details of this spectrum, the compressibility, and  the 
optical conductivity should be disclosed in the quantum treatment. 
\subsection{Massive Thirring model}
 To solve eq.(\ref{phasecharge}) quantum mechanically, we describe
 the charge degree of freedom in terms of spinless Fermion
\cite{emery1,emery2,giamarchi,emery3,mori}
instead of Boson. Eq.(\ref{phasecharge}) is mapped onto 
the following Hamiltonian (See Appendix A),
\begin{eqnarray}
H_{\rho,F}  
 & = & v_c\int {\rm d}x (\Psi^{\dag}(x) (- {\rm i}\partial\tau_{3})\Psi(x))+ \frac{v_c q_0}{2} \Psi^{\dag}(x)\Psi(x)\nonumber\\
 & + & V\int {\rm d}x \Psi^{\dag}(x) \tau_{1} \Psi(x)\nonumber\\
 & + & \frac{W}{2/\pi}\int {\rm d}x[ (\Psi^{\dag}(x)  \Psi(x))^2
-(\Psi^{\dag}(x) \tau_{1} \Psi(x))^2],
\label{spless}
\end{eqnarray}
where
\beqa
\Psi
&=&
\left(\begin{array}{c}
                \psi_{1}(x) \\ \psi_{2}(x)
                \end{array}\right), \nonumber\\
v_c\nonumber
& = &\pi A_{\rho}+C_{\rho}\frac{1}{\pi},\\
V\nonumber
& = & B_{\rho}(\pi\alpha),\\
W\nonumber
& = &\pi A_{\rho}-C_{\rho}\frac{1}{\pi},
\label{mapping2}
\eeqa
and $\tau_j$, (${\it j}$=0,1,2,3) are Pauli matrices. The second term, which originates from
  the Umklapp scattering process, creates the charge gap (Mott-Hubbard gap).
This field operator, $\Psi(x)$, is identified as a charge particle because  
$\Psi(x)$ creates a DC in $\theta(x)$ as follows\cite{mandelstam,emery2},
\beq
[\theta(x), \psi_j^{\dag}(y)]=
{\frac{\pi}{2} {\rm sgn}}(x-y)\psi_j^{\dag}(y) \hspace{1cm}(j=1,2).\nonumber\\
\eeq
As mentioned in the preceding section (see eq.(\ref{soliton})), this DC  
corresponds to a charge particle.
 In the following discussion, we will set $W=0$
and then eq.(\ref{spless}) represents a system of free Fermions.
This particular choice, $W=0$, is a good approximation near the Mott transition
as has been noted by Giamarchi.\cite{giamarchi} 
The reason is as follows; if $W=0$ we see
\beqa
A_{\rho}\nonumber
&=&\frac{v_{\rho}}{4\pi K_{\rho}},\\
C_{\rho}\nonumber
&=&\pi v_{\rho}K_{\rho},
\eeqa
therefore, 
\beq
K_{\rho}=\frac{1}{2},\nonumber
\eeq
 where $v_{\rho}$ is the velocity of charge 
excitation and $K_{\rho}$ is the critical exponent for the density 
correlation function. On the other hand, it is known that 
both quantities, $v_{\rho}$ and $K_{\rho}$, are generally determined by the 
 degree of the filling of band and the coupling constant. 
Especially, $K_{\rho}$ approaches $1/2$ near the half-filling for any values of 
coupling constant.\cite{kawakami,korepin,schulz}
It is noted that  $K_{\rho}$ also becomes to $1/2$ toward 
the infinite repulsive 
interaction for any filling as noted by Emery.\cite{emery2}
\subsection{Energy Spectra and Compressibility}
   With $W=0$, we can solve eq.(\ref{spless}) exactly and 
get a simple physical picture about the charge degree
of freedom near the Mott transition.\cite{emery2,giamarchi,emery3,mori} 
First, the energy dispersion is 
  \beq
E_{-}(k)=\frac{v_cq_0}{2}-\sqrt{v_c^2k^2+V^2},\nonumber\\
\label{lowband}
\eeq
\beq
E_{+}(k)=\frac{v_cq_0}{2}+\sqrt{v_c^2k^2+V^2}.
\label{upperband}
\eeq
The band structure relative to the chemical potential is shown schematically 
in Fig.3. 
The actual carrier density (doped hole density) away from half-filling, $\delta$ , 
is determined by the range of $k$ which satisfies $E_{-}(k)\geq0$.
For $q_0>q_c\equiv 2V/v_c$ , $\delta$ is determined by,
\beq
\delta=\frac{v_c}{2\pi V}\sqrt{q_0^2-q_c^2},
\label{dopingrate}
\eeq
 and $\delta=0$ for $q_0<q_c$.
This relation determines the carrier density as a function of the chemical potential, $q_0$. 
Consequently we immediately see  that 
the compressibility, $\kappa=\partial n/\partial \mu$, diverges inversely proportional
to the doping rate\cite{giamarchi,mori} as found in the study by Usuki, Kawakami 
and Okiji based on the Bethe Ansatz\cite{usuki},
because of the particular energy dependence of the density of states at the 
band edge in one dimension.
\section{Charge Excitation and Conductivity in the Clean System}
 We will first calculate the density-density correlation function and 
next the conductivity and Drude weight by use of the Nambu formalism 
(See Appendix B).
\subsection{Charge Excitation Spectrum}
  The charge excitation spectrum is given by the imaginary part of the retarded 
density-density correlation function, ${\rm Im
N}^R(q,\omega)$.
Since we are interested in the spatially slowly varying  part ($q\ll2k_0$), 
the charge density is given by 
$n(x,t)=e\Psi^{\dag}(x,t) \Psi(x,t)$. 
After straightforward calculation, we obtain the following result,
\beqa
{\rm Im N}^R(q,\omega)
&=&
\frac{2 e^{2}}{\hbar} \int \frac{{\rm d}k}{2 \pi}\int \frac{{\rm d}\epsilon}{2 \pi} 
(f(\epsilon)-f(\epsilon+\omega))\rm{Tr} [\tau_0{\rm Im \hat{G}}^R(k+q,\epsilon +\omega)
\tau_0{\rm Im \hat{G}}^R(k,\epsilon)]\nonumber\\
&=&
\frac{2 e^{2}}{\hbar} \int \frac{{\rm d}k}{2 \pi} 
[(v_{k}v_{k+q}+u_{k}u_{k+q}+\frac{V^2}{2 E_{k}E_{k+q}})
\delta(\omega+E_{k+q}-E_{k})\nonumber\\
&+&
(v_{k}u_{k+q}+u_{k}v_{k+q}-\frac{V^2}{2 E_{k}E_{k+q}})
\delta(\omega-E_{k+q}-E_{k})],
\label{nqw}
\eeqa
where $\tau_0$ is the unit matrix and 
\begin{eqnarray}
   u_k = \frac{1}{2} (1 + \frac{k}{E_{k}}), \nonumber\\
   v_k = \frac{1}{2} (1 -\frac{k}{E_{k}}), \nonumber\\
   E_{k}=\sqrt{v_c^2 k^2 +V^2}.
\end{eqnarray}
In the last equation, the k-integration is carried out for 
$v_{c}q_0/2<E_{k}<v_{c}q_0/2+\omega$, 
$f(\epsilon)$ is the Fermi distribution function and 
${\rm \hat{G}}(k,{\rm i}\epsilon)$ is Fourier transform of Green function defined as follows, 
\beqa
{\rm \hat{G}}(x,\tau)
&=&
-\langle 
T_{\tau}
\left(
\begin{array}{cc}
\psi_{1}(x,\tau)  \psi_{1}^{\dag}(0,0) & \psi_{2}(x,\tau)  \psi_{1}^{\dag}(0,0)\\
\psi_{1}(x,\tau)  \psi_{2}^{\dag}(0,0) & \psi_{2}(x,\tau)  \psi_{2}^{\dag}(0,0)
\end{array}
\right )
\rangle.\nonumber
\eeqa
The detailed explanation of the Green function is given in Appendix B.
From eq.(\ref{nqw}), we see that the charge excitation exists in the following 
regions;\\
1.Acoustic Excitation:
\beqa
(v_{c}q)^2+4V^2-\omega^2&\geq&0, \nonumber\\
(v_{c}q)^2-\omega^2&>&0, \nonumber\\
v_{c}q_{0}-\omega \leq  \nonumber
(v_{c}q)\sqrt{\frac{(v_{c}q)^2+4V^2-\omega^2}{(v_{c}q)^2-\omega^2}}&\leq& \nonumber
v_{c}q_{0}+\omega.
\eeqa
2.Optical Excitation:
\beqa
(v_{c}q)^2+4V^2-\omega^2&\leq&0, \nonumber\\
(v_{c}q)^2-\omega^2&<&0,\nonumber\\
\left\{  v_{c}q_{0}-\omega \leq \right.\nonumber
v_{c}q\sqrt{\frac{(v_{c}q)^2+4V^2-\omega^2}{(v_{c}q)^2-\omega^2}}&\leq& \nonumber
v_{c}q_{0}+\omega \nonumber\\
{\rm or}\nonumber\\
-(v_{c}q_{0}-\omega) \geq \nonumber
v_{c}q\sqrt{\frac{(v_{c}q)^2+4V^2-\omega^2}{(v_{c}q)^2-\omega^2}}&\geq& \nonumber
\left. -(v_{c}q_{0}+\omega )  \right\}.
\eeqa
These regions are plotted for several choices of doping rate in Fig.4 and 
actual magnitude of the spectral weight, ${\rm Im N}^R(q,\omega)$, in these 
regions are shown in Fig.5.\cite{mori}
The existence of the acoustic mode of excitation spectrum has also been obtained 
from a quantum Monte Carlo simulation using the maximum entropy method for 
the 1D Hubbard model away from half-filling.\cite{preuss} 
In the metallic state (far away from half-filling), the acoustic mode, which 
corresponds to  the charge excitation in the Tomonaga-Luttinger liquid, dominates
 spectral weight especially around the small wave numbers. 
Thus it is reasonable to consider that the inter-band excitation across 
the Mott-Hubbard gap does not contribute to the density-density correlation 
function in the limit of long wave length. 
Near the Mott transition, however, such region where the acoustic mode is 
dominant becomes small and the optical mode is dominant elsewhere. 
This implies that, in the real space, the density-density correlation function has a 
characteristic length scale for the crossover behavior. The similar crossover 
behavior is numerically found in the spin-spin correlation function.\cite{imada}  
As  the Mott insulator is approached, 
the characteristic length scale becomes large, above which  the correlation function follows 
its asymptotic form derived by the conformal field theory. 
In the limit of Mott insulator, this length scale diverges and the conformal field 
theory breaks down.  Correspondingly, the spectral weight shifts from the 
acoustic mode to the optical mode and the metallic phase crossover 
to the insulator phase. The Mott transition is characterized by this crossover,\cite{mori} 
which has not been so far found out by any theories.  
The  location in $q$-space of this crossover about the dominant spectral weight 
corresponds to that of  the excitation spectrum in the classical approximation
 as shown in Fig.2.
\subsection{Conductivity and Drude Weight}
  Next the conductivity\cite{giamarchi,giamarchi2,giamarchi3} and Drude weight\cite{giamarchi,emery3,mori} 
will be calculated by 
the imaginary part of retarded current-current correlation function, 
${\rm Im K}^R(\omega)$, 
which is evaluated as follows.
By noting that the current density is 
$j(x,t)=e v_{c}\Psi^{\dag}(x,t)\tau_{3} \Psi(x,t)$ because
of the continuity equation,
we obtain
\beqa
\lim_{q\rightarrow 0}{\rm Im K}^{R}(q,\omega)
&=&
\frac{2 e^{2}v_{c}^2}{\hbar} 
\int \frac{{\rm d}k}{2 \pi}\int \frac{{\rm d}\epsilon}{2 \pi} 
(f(\epsilon)-f(\epsilon+\omega))
\lim_{q\rightarrow 0}{\rm Tr [Im \hat{G}}^R(k+q,\epsilon +\omega)
\tau_{3} {\rm Im\hat{G}}^R(k,\epsilon)\tau_{3}]\nonumber\\
&=&
\frac{2 e^{2}v_{c}^2}{\hbar} \int \frac{{\rm d}k}{2 \pi} 
\lim_{q\rightarrow 0}
[(v_{k}v_{k+q}+u_{k}u_{k+q}-\frac{V^2}{2 E_{k}E_{k+q}})
\delta(\omega+E_{k+q}-E_{k})\nonumber\\
&+&
(v_{k}u_{k+q}+u_{k}v_{k+q}+\frac{V^2}{2 E_{k}E_{k+q}})
\delta(\omega-E_{k+q}-E_{k})].
\label{currentcorrelation}
\eeqa
In the last equation, the k-integration is carried out for 
$v_{c}q_0/2< E_{k}< v_{c}q_0/2+\omega$.\\
Near the Mott transition, the Drude weight, $D_c$, which is defined as, 
${\rm Re \sigma}(\omega)=D_c {\rm \delta}(\omega)$, is calculated as follows 
(see also Appendix C), 
\beq
{\rm Re}\sigma(\omega)
=
{\rm Re}\frac{K^{R}(\omega) - 2 v_c e^2/h}
{{\rm i} \omega - \eta},\nonumber\\
\eeq
therefore,
\beqa
 D_c &=&
\frac{e^2 v_{c}}{\hbar}\frac{\delta}{q_0/2},\nonumber\\
& \sim &
\frac{e^2 v_{c}}{\hbar}\frac{\delta}{V}.
\label{drudeweight}
\eeqa
This fact, that $D_c$ is proportional to the concentration of the doped holes 
near the Mott transition, 
 is consistent with the exact result obtained by Shastry and Sutherland by 
use of the Bethe Ansatz\cite{shastry}
 and has also been noted by Giamarchi\cite{giamarchi} and Emery.\cite{emery3}
It is interesting to notice that the doping dependence of $D_c$ is different from 
that of Tomonaga-Luttinger liquid.
In the Tomonaga-Luttinger regime, the $D_c$ is determined as, 
$2 e^2 v_{c}/\hbar$, by the electron 
(particle) number density  which is expressed through  $v_{c}$.\cite{fukuyama}
This result is obtained by the Kubo formula by using the anomalous commutation relation 
which is inherent to the one dimensional systems with linear dispersion.
\cite{haldane3}
The different dependence on the carrier concentration of $D_c$ between in the 
Tomonaga-Luttinger regime and near the Mott insulator results from the roles played 
by the Umklapp scattering process.
 Thus we can understand  that the Umklapp  
scattering process plays crucial roles near the Mott transition. 
\section{Impurity Effect}
 So far we confined ourselves to the clean system. We will in the following consider 
the effects of disorder represented by the impurity scattering.  
 We first  explain how to introduce and treat impurity scattering, and then 
we study the effects of impurity scattering on the density 
of states, which determines the compressibility and the conductivity. 
We will employ the self-consistent Born approximation and ignore the possible
consequence of the Anderson localization or the bound states in the Mott-Hubbard
gap in order to understand overall features in the present study.
\subsection{Impurity Hamiltonian and Self-Energy}
  The impurity potential, $u(x)$, couples to the charge density, $\rho(x)$, as follows,
\beqa
H_{imp}
&=&\sum_i\int{\rm d}x u(x-R_i)\rho(x)
\label{impurityscattering}
\eeqa
where $\rho(x)$ is expressed as follows,
\beqa
\rho(x)
&=&\psi_1^{\dag}(x)\psi_1(x) +\psi_2^{\dag}(x)\psi_2(x)\nonumber\\
&+&e^{{\rm -i}\lambda x}\psi_1^{\dag}(x)\psi_2(x)+
e^{{\rm i}\lambda x}\psi_2^{\dag}(x)\psi_1(x). 
\eeqa
The oscillating part of $\rho(x)$, which is characterized by the 
constant, $\lambda$, originates in the density fluctuation across 
the Fermi surface of spinless fermion. 
We suppose the impurity potential to be the $\delta$-function, 
$u(x)=u \delta(x)$. In this impurity scattering Hamiltonian, we 
ignored the coupling between the charge and spin excitation through the 
$2k_F$-oscillating part of charge density.\cite{suzumura-cardy}

In the second order self-consistent Born approximation, 
the self-energy, which is diagrammatically 
shown in the Fig.6, is given as follows; 
\beqa
\hat{\Sigma}(z)
& = &
\left(
\begin{array}{cc}
\sigma(z) & \Delta^{\dag}(z)\\
\Delta(z)& \sigma(z)
\end{array}
\right )\nonumber\\
&=&
n_{i}u^2\int \frac{\rm dk}{2 \pi} 
\left(
\begin{array}{cc}
{\bf g}_1(k,z) +{\bf g}_2(k,z) & {\bf f}^{\dag}(k,z)\\
{\bf f} (k,z)& {\bf g}_1(k,z) +{\bf g}_2(k,z)
\end{array}
\right ),
\label{selfenergy}
\eeqa
where 
\beqa
\sigma(z)
&=& 
{\rm -i}\frac{n_{i}u^2}{v_c}\frac{\tilde{z}}{\sqrt{\tilde{z}^2-\tilde{V}^2}}\nonumber\\
\Delta(z)=\Delta^{\dag}(z)
&=&
{\rm -i}\frac{n_{i}u^2}{2v_c}\frac{\tilde{V}}{\sqrt{\tilde{z}^2-\tilde{V}^2}},
\label{selfenergy2}
\eeqa
which are easily obtained after k-integration. For details of $\tilde{z}, \tilde{V}$, 
see Appendix B.   
The self-consistent equations eq.(\ref{selfenergy}) and  eq.(\ref{selfenergy2}) are  
reduced to the following equations,
\beqa
\tilde{z}=z-\sigma(z)
&=&
z+{\rm i}\frac{n_{i}u^2}{v_c}\frac{\tilde{z}}{\sqrt{\tilde{z}^2-\tilde{V}^2}},\nonumber\\
\tilde{V}=V+\Delta(z)
&=&
V-{\rm i}\frac{n_{i}u^2}{2v_c}\frac{\tilde{V}}
{\sqrt{\tilde{z}^2-\tilde{V}^2}}.
\label{selfconsistenteq}
\eeqa
Eq.(\ref{selfconsistenteq}) is rewritten as 
\beq
\chi^4-2y\chi^3+(y^2+\zeta^2-1)\chi^2+2y\chi-y^2=0
\label{selfconsistenteq2}
\eeq
where $\zeta=3n_{i}u^2/2v_c V$ , $\chi=\tilde{z}/\tilde{V}$ and $y=z/V$.
\subsection{Density of States and Compressibility}
  By using the solution of eq.(\ref{selfconsistenteq2}), the density of states,
$\cal{D}(\epsilon)$, is 
calculated as follows,
\beqa
\cal{D}(\epsilon)
&=&
-\frac{1}{\pi} \int\frac{{\rm d}k}{2\pi}{\rm Im} {\rm Tr}[\hat{\bf G}^R(k,z)],\nonumber\\
&=&
\frac{1}{\pi v_c} {\rm Re} \frac{\chi}{\sqrt{\chi^2-1}},\nonumber\\
&=&
\frac{1}{\pi v_c \zeta} {\rm Im}(\chi).
\eeqa
$\cal{D}(\epsilon)$ is plotted in Fig.7 for several choices of dimensionless  
parameter, $\zeta$, representing the strength of impurity scattering.
It should be noted that, in the presence of impurity scattering, the Mott-Hubbard 
gap, which is considered to be  the energy gap in the density of states of spinless 
Fermion,  exists for $0< \zeta< 1$. In the following, we will use the term 'weak 
disordered system' to refer to the system for $0< \zeta< 1$. On the other hand, for 
$\zeta \geq 1$, the gap disappears but the density of states has a cusp at 
$\epsilon=0$. 

In addition to the change of Mott-Hubbard gap, the $\cal{D}(\epsilon)$ in the disordered 
systems does not diverge at the band edge in contrast to the $\cal{D}(\epsilon)$ 
in the clean system. 
This qualitative change of density of states directly affects  the 
compressibility, $\kappa=\partial n/\partial \mu$; In the clean system $\kappa$ 
diverges, but in the weak disordered system,  
it smoothly vanishes as $\delta^{2/3}$ 
as is shown in Fig.8.   
 Though the value of the  critical exponent $2/3$ should not be taken literally in view of 
the approximation employed, this drastic qualitative difference between the clean 
and disordered systems is to be noted. For $\zeta \geq 1$, 
$\kappa$ approaches a finite constant.
\subsection{Charge Excitation Spectrum  in the Disordered System}
   In the calculation of the density-density correlation function, ${\rm N}(q,\omega)$, 
 we need the vertex 
correction in addition to the self-energy correction.
 The vertex corrections to the density vertex, $\hat{\Gamma}_N(q,\omega)$, 
is the solution of following Dyson equation,
\beqa
\hat{\Gamma}_N(q,{\rm i}\omega)
&=&
\tau_0+\frac{2\zeta}{3 V} \int\frac{{\rm d}k}{2\pi} \left[ \hat{\bf G}(k+q,z+{\rm i}\omega)
\tau_0\hat{\bf G}(k,z) \right. \nonumber\\
&+&
\tau^+ \hat{\bf G}(k+q,z+{\rm i}\omega)\tau_0
\hat{\bf G} (k,z)\tau^-\nonumber\\
&+&
\tau^- \hat{\bf G} (k+q,z+{\rm i}\omega)\tau_0
\hat{\bf G} (k,z)\tau^+\nonumber\\
&+&
\hat{\bf G} (k+q,z+{\rm i}\omega)
\hat{\Gamma}_N(q,{\rm i}\omega)
\hat{\bf G} (k,z)\nonumber\\
&+&
\tau^+ \hat{\bf G} (k+q,z+{\rm i}\omega)
\hat{\Gamma}_N(q,{\rm i}\omega)
\hat{\bf G} (k,z)\tau^-\nonumber\\
&+&
\left.
\tau^- \hat{\bf G} (k+q,z+{\rm i}\omega)
\hat{\Gamma}_N(q,{\rm i}\omega)
\hat{\bf G} (k,z)\tau^+\right],
\label{densityvertex}
\eeqa
where $\tau^{\pm}=(\tau_1 \pm {\rm i}\tau_2 )/2$ and $\tau_0$ is the unit matrix. 
${\rm N}(q,{\rm i}\omega)$  is calculated by using the solution of the above equation 
as follows, 
\beq
{\rm N}(q,{\rm i}\omega)=\frac{{\rm e}^2}{h}({\it -T})\sum_z\int\frac{{\rm d}k}{2\pi}
{\rm Tr}\left[\tau_0\hat{\bf G} (k+q,z+{\rm i}\omega)\hat{\Gamma}_N(q,{\rm i}\omega)
\hat{\bf G} (k,z)\right]. 
\eeq
$\rm Im N(q,\omega)$ are plotted in Fig.9 for several values of 
$\delta$, $\zeta$ and $q$, by solving the above equation numerically. \\
In the weak disordered systems, there also exist the acoustic excitation and the optical 
excitation as in the clean system,  
though these excitations are damped.
Hence, the crossover feature of spectral weight in $q$ and $\omega$-space 
survives as far as the impurity scattering is not so strong as seen in Fig.9 for $\zeta=0.1$ and $\delta=0.3$. 
The doping dependence of spectral weight 
is shown in Fig.9 for  $\zeta=0.1$ and $q=2$, where it is seen that 
the dominant excitation smoothly
changes from the acoustic (optical) excitation to the optical (acoustic) excitation 
as the doping rate is varied.  
Therefore the crossover behavior in the vicinity of the Mott transition
 is visible not only in the clean system but is present also in the weak 
disordered systems as well. 
\subsection{Optical Conductivity and Drude Weight in the Disordered System}
   In the calculation of the conductivity, we need the vertex correction for the 
current vertex, $\hat{\Gamma}_K(0,\omega)$, again 
similar to that of density-density correlation function; \ie, 
insert $\tau_3$ between Green functions for the first, second and third term 
in eq.(\ref{densityvertex})  and take the limit of $q\rightarrow0$.
\beqa
\hat{\Gamma}_K({\rm i}\omega)
&=&
\tau_3+\frac{2\zeta}{3 V} \int\frac{{\rm d}k}{2\pi} \left[ \hat{\bf G} (k,z+{\rm i}\omega) \tau_3
\hat{\bf G} (k,z) \right. \nonumber\\
&+&
\tau^+ \hat{\bf G} (k,z+{\rm i}\omega) \tau_3
\hat{\bf G} (k,z)\tau^-\nonumber\\
&+&
\tau^- \hat{\bf G} (k,z+{\rm i}\omega) \tau_3
\hat{\bf G} (k,z)\tau^+\nonumber\\
&+&
\hat{\bf G} (k+q,z+{\rm i}\omega)
\hat{\Gamma}_K({\rm i}\omega)
\hat{\bf G} (k,z)\nonumber\\
&+&
\tau^+ \hat{\bf G} (k+q,z+{\rm i}\omega)
\hat{\Gamma}_K({\rm i}\omega)
\hat{\bf G} (k,z)\tau^-\nonumber\\
&+&
\left.
\tau^- \hat{\bf G} (k,z+{\rm i}\omega)
\hat{\Gamma}_K({\rm i}\omega)
\hat{\bf G} (k,z)\tau^+\right].
\label{currentvertex}
\eeqa
The current-current correlation function, $K({\rm i}\omega)$, with vertex correction is 
calculated as follows, 
\beq
{\rm K}({\rm i}\omega)=\frac{({\rm e}v_c)^2}{h}({\it -T})\sum_z
\int\frac{{\rm d}k}{2\pi}
{\rm Tr}\left[\tau_3\hat{\bf G} (k,z+{\rm i}\omega)\hat{\Gamma}_K({\rm i}\omega)
\hat{\bf G} (k,z)\right]. 
\eeq
The optical conductivity is  obtained by solving the above equation numerically. 
In Fig.10, the $\omega$-dependence of optical conductivity in the case of 
$\delta$=0.01 is shown  for several choices of $\zeta$=0.0, 0.01, 0.1, 0.3.
There exist optical absorption across the Mott-Hubbard gap as well as
 the Drude tail. Because of the impurity scattering, both spectra become broad 
and difficult to distinguish from each another if the impurity scattering becomes strong.
 An example of the doping dependence of the optical conductivity in the weak 
disordered systems is plotted for $\zeta=0.1$ and some choices of $\delta$ in Fig.11.
 In this figure, the Drude weight seems to behave  
 similarly  to that of clean system as a function of doping rate, 
\ie, smoothly approaches to zero. It becomes clear by comparing the integrated 
weight of both, Drude and optical,  that the missing weight of Drude peak shifts to 
the optical absorption and 
the metallic state merges to the insulator phase. These general features of optical 
conductivity is plausible but it has not been explicitly demonstrated so far. 
The crossover feature near the Mott transition 
appears as the smooth shift of weight in the optical conductivity.
\section{Conclusion and Discussion}
  In the present study, we investigated the excitation spectrum of density-density 
correlation function and the optical 
conductivity near the Mott transition in one-dimensional electron system
in the presence of disorder.
Those quantities were calculated based on the massive Thirring model for the
 spinless Fermion.
In the Mott insulator, \ie, at half-filling, the Mott-Hubbard gap was seen to 
originate from the Umklapp scattering process as noted by 
Emery.\cite{emery1,emery2,emery3}
Even in the doped Mott insulator, \ie, away from half-filling, this process
 played the important role  as indicated by Giamarchi.\cite{giamarchi}
First, we calculated the static limit of the density-density correlation function and 
the conductivity, \ie, compressibility\cite{mori,giamarchi} and Drude weight.\cite{giamarchi,emery3,mori} In the clean system, 
the doping dependence of both quantities were same as the results deduced by the 
exact solution.\cite{usuki}  Although the Drude weight 
did not qualitatively change its behavior also in the weak disordered systems, 
the compressibility even in the weak disordered systems behaved  completely 
different way from that in the clean system. Next, we studied the dynamical 
properties. In both the clean and weak disordered systems, 
the crossover behavior existed in the spectra of density-density correlation 
function; in the region of small wave number, the acoustic mode was dominant 
while in the large wave number region, the optical mode was weighted. 
 By decreasing the doping rate, the region, in which the acoustic mode was dominant, 
became small and its spectral weight shifted to that of optical mode. 
We would like to emphasize that the Mott transition was characterized by the 
crossover feature rather than drastic phase transition. 
The smooth shift of spectral weight from the Drude weight to the optical 
absorption is present in the optical conductivity 
in both the clean and weak disordered systems. If the impurity scattering becomes 
strong and the effect of localization becomes important,\cite{suzumurafukuyama} the optical conductivity 
would vanish as approaching $\omega=0$ and will show an characteristic features of
 the pinning frequency.\cite{fl}
\section*{Acknowledgements}
M.M. would like to thank Hiroshi Kohno and Hideaki Maebashi for their valuable
 discussions. Thanks are due to Professor T. Giamarchi and Professor V. J. Emery 
for valuable comments. This work was financially supported by a Grant-in-Aid for Scientific 
Research on Priority Area "Anomalous Metallic State near the Mott 
Transition" (07237102) from the Ministry of Education, Science, Sports and Culture.

\appendix
\section{Mapping from phase Hamiltonian to massive Thirring model}
The equivalence between the two Hamiltonians eq.(\ref{phasecharge}) and 
(\ref{spless}) is proved by the bosonization method.
To begin with, we define the boson operators, 
$\theta_j(x)$ $(j=1,2)$, in terms of the spinless Fermions $\psi_j(x)$ $(j=1,2)$ 
as follows,
\begin{equation}
\theta(x)\equiv\frac{1}{2}(\theta_{1}(x)+\theta_{2}(x)),
\label{chargeboson}
\end{equation}
\begin{equation}
\theta_{j}\equiv {\rm i}\sum_{q}\frac{2\pi}{Lq}\exp({\rm -i}qx-\frac{1}{2}\alpha|q|)
\rho_{j}(q), \; (j=1,2),
\label{boson}
\end{equation}
\beqa
\rho_{j}(q)&\equiv&\sum_{k}c_{j,k+q}^{\dag}c_{j,k}, \; (q > 0), \nonumber\\ 
\rho_{j}(-q)&\equiv&\sum_{k}c_{j,k}^{\dag}c_{j,k+q}, \; (q > 0), 
\label{splessdensityop}
\eeqa
where 
\beq
[\rho_{1}(-q), \rho_{1}(q')]=[\rho_{2}(q), \rho_{2}(-q')]=\frac{Lq}{2\pi}\delta_{q,q'}.\nonumber
\label{splessdenscommu}
\eeq
The spinless Fermions correspond to those  Boson operators 
as follows,
\begin{equation}
\psi_i(x)=\frac{1}{\sqrt{2\pi\alpha}}\exp(\pm {\rm i}\theta_i(x)), \; (+\:for\:i=1,-\:for\:i=2).
\label{bosonization}
\end{equation}
The Hamiltonian eq.(\ref{phasecharge}) is described in terms of the density operators 
as follows,
\beqa
\int {\rm d}x  (\nabla\theta(x))^2
&=&\pi\int_{q>0} {\rm d}q \left[\rho_1(q)\rho_1(-q)+\rho_2(-q)\rho_2(q)+
\rho_1(q)\rho_2(-q)+\rho_1(-q)\rho_2(q)\right],\nonumber\\
\int {\rm d}x  (P(x))^2
&=&\frac{1}{\pi}\int_{q>0} {\rm d}q \left[\rho_1(q)\rho_1(-q)+\rho_2(-q)\rho_2(q)-
\rho_1(q)\rho_2(-q)-\rho_1(-q)\rho_2(q)\right],\nonumber
\eeqa
thus,
\beqa
A_{\rho}\int {\rm d}x  (\nabla\theta(x))^2+C_{\rho}\int {\rm d}x  (P(x))^2
\nonumber
&=&\left(\pi A_{\rho}+\frac{C_{\rho}}{\pi}\right)
\int_{q>0} {\rm d}q \left[\rho_1(q)\rho_1(-q)+\rho_2(-q)\rho_2(q)\right]\nonumber\\
&+&
\left(\pi A_{\rho}-\frac{C_{\rho}}{\pi}\right)
\int_{q>0} {\rm d}q \left[\rho_1(q)\rho_2(-q)+\rho_1(-q)\rho_2(q)\right].
\label{densityphase}
\eeqa
Next, the first and third term of eq.(\ref{spless}) are also written in terms of 
the density operators as follows,
\beqa
H_{\rho,D}^{(1)}
&=&v_c\int_{q>0} {\rm d}q\left[\rho_1(q)\rho_1(-q)+\rho_2(-q)\rho_2(q)\right],
\nonumber\\
H_{\rho,D}^{(3)}
&=&W\int_{q>0} {\rm d}q \left[\rho_1(q)\rho_2(-q)+\rho_1(-q)\rho_2(q)\right]
\label{densityspless}
\eeqa
which are proved by the following equations, \eg, 
\subbeqa
\left[H_{\rho,F}^{(1)}, \rho_{1,\sigma}(q)\right]
&=&v_cq\rho_{1,\sigma}(q),\\
\left[H_{\rho,D}^{(1)}, \rho_{1,\sigma}(q)\right]
&=&v_cq\rho_{1,\sigma}(q),
\subeeqa
for the first term of eq.(\ref{spless}). 
By comparing eq.(\ref{densityphase}) and eq.(\ref{densityspless}), we see 
the following relations,
\beqa
v_c\nonumber
& = &\pi A_{\rho}+C_{\rho}\frac{1}{\pi},\\
W\nonumber
& = &\pi A_{\rho}-C_{\rho}\frac{1}{\pi},
\eeqa
$B_{\rho}$ term is easily derived by inserting eq.(\ref{bosonization}) into the second
term of eq.(\ref{spless}).
\section{Definition of Green Function}
  In the Nambu formalism for the charge degree of freedom near the Mott transition, 
the Green function, $\hat{G}(x,\tau)$,  is defined as  $2\times2$ matrix,  
\beqa
\hat{G}(x,\tau)
&=&
-\langle 
T_{\tau}
\left(
\begin{array}{cc}
\psi_{1}(x,\tau)  \psi_{1}^{\dag}(0,0) & \psi_{2}(x,\tau)  \psi_{1}^{\dag}(0,0)\\
\psi_{1}(x,\tau)  \psi_{2}^{\dag}(0,0) & \psi_{2}(x,\tau)  \psi_{2}^{\dag}(0,0)
\end{array}
\right )
\rangle,\nonumber\\
\eeqa
where $T_{\tau}$ means the chronological order in the imaginary time.
In the clean system, after Fourier transformation for the system without the interaction, $W=0$, 
\beqa
\hat{G}^0 (k,z)
&=&
\left( 
\begin{array}{cc}
g_1^0(k , z) & f^{\dag,0}(k , z)\\
f^0(k , z)     &  g_2^0(k , z)
\end{array}
\right),\nonumber\\
& = &
\left(
\begin{array}{cc}
z - v_ck & -V\\
-V & z+v_ck
\end{array}
\right )^{-1},
\eeqa
where $ z=i \epsilon- v_c q_0/2$.\\
In the disordered systems, full Green function, $\hat{\bf G}_ (k,z)$, including the 
self-energy correction is calculated as 
follows,
\beqa
\hat{\bf G}(k,z)
& = &
\left(
\hat{G}^0 (k,z)^{-1}-\hat{\Sigma}(z)
\right )^{-1},\nonumber\\
& = &
\left(
\begin{array}{cc}
z - v_ck -\sigma(z) & -V -\Delta^{\dag}(z)\\
-V -\Delta(z)& z+v_ck-\sigma(z)
\end{array}
\right )^{-1}.
\eeqa
We take the full
Green function as follows to determine the 
self-energy (or Green function),
\beqa
\hat{\bf G}(k,z)
&=&
\left(
\begin{array}{cc}
{\bf g}_1(k,z) & {\bf f}^{\dag}(k,z)\\
{\bf f} (k,z)& {\bf g}_2(k,z)
\end{array}
\right ),\nonumber\\
& = &
\left(
\begin{array}{cc}
\tilde{z} - v_ck & -\tilde{V}^{\dag}\\
-\tilde{V}& \tilde{z}+v_ck
\end{array}
\right )^{-1}.
\label{fullgreen}
\eeqa
We can derive the self-consistent equation eq.(\ref{selfconsistenteq}) by 
substituting eq.(\ref{fullgreen}) to eq.(\ref{selfenergy}).
\section{Calculation of Drude Weight near the Mott Transition}
    An uniform electric field, $\rm E_x(t)$, along the one dimensional system has such a
relation, $\rm E_x(t)=\partial_x A_0(x,t) - \partial_t A_1(x,t)$,
to the vector potential, $\mbox{\boldmath$A$}(x,t)=(A_0(x,t),A_1(x,t))$.
To make our discussion simple, we introduce external field as follows, 
$\mbox{\boldmath$A$}(x,t)=(x\rm E_x(t),0)$. 
The perturbation term by the external field, $\cal{H}_{\rm ext}$, is introduced as 
ordinary way,
\beq
{\cal H}_{\rm ext}=-\int{\rm d}x {\rm n}(x)A_0(x,t).\nonumber
\eeq
On these grounds, the conductivity, $\sigma(\omega)$, is calculated as follows,
\beqa
\sigma(\omega)
&=&
\frac{\rm i}{\hbar}\int^{\infty}_{0}{\rm d}s\int^{\infty}_{-\infty}{\rm d}y
{\rm e}^{{\rm i} \omega s} \langle \left[ {\rm J}(s,0),{\rm n}(0,y) \right] \rangle 
y,
\nonumber\\
&=&
\frac{\rm i}{\hbar}\int^{\infty}_{-\infty}{\rm d}y\frac{-1}{{\rm i} \omega-\eta}
\langle \left[ {\rm J}(0,0),{\rm n}(0,y) \right] \rangle y\nonumber\\
&+&
\frac{\rm i}{\hbar}\int^{\infty}_{0}{\rm d}s\frac{{\rm e}^{({\rm i} \omega-\eta)s}}
{{\rm i} \omega-\eta}\int^{\infty}_{-\infty} {\rm d}y\langle \left[ {\rm J}(0,s),
{\rm J}(0,y)\right] \rangle.
\label{conductivity}
\eeqa
where the translational invariance in space coordinate and the continuity 
equation are noted. \\
In eq.(\ref{conductivity}), the first term in the last equation is estimated as,
\beq
\frac{\rm i}{\hbar}\int^{\infty}_{-\infty}{\rm d}y\frac{-1}{{\rm i}\omega-\eta}
\langle \left[ {\rm J}(0,0),{\rm n}(0,y) \right] \rangle y
=\frac{2 e^2 v_{c}}{h}\frac{-1}{{\rm i} \omega - \eta},\nonumber
\eeq
by using the anomalous commutation relation,
\beq
\left [ {\rm J}(x),{\rm n}(y) \right]=\frac{{\rm i}v_{c}}{\pi}\partial_x \delta(x-y).
\label{anomalous}
\eeq
 As a result, the real part of conductivity is calculated as follows,
\beqa
{\rm Re}\sigma(\omega)
&=&
{\rm Re}\frac{K^{R}(\omega) - 2 v_c e^2/h}
{\rm{i} \omega - \eta},\nonumber\\
&=&
\left\{ {\rm Re}K^{R}(\omega) - \frac{2 v_c e^2}{h} \right\} (-\pi)\delta(\omega)
+
\frac{\rm{Im}K^{R}(\omega)}{\omega}.
\eeqa
Without the Umklapp scattering process in the last equation, the terms 
corresponding to the $K^{R}(\omega)$ is zero and the Drude weight
takes the constant value, $2 e^2 v_{c}/h$.

\newpage
${\Large {\bf Figure  Caption}}$\\
Fig.1 The classical solution is plotted for $m$=1 and $k^2$=0.5, 0.9, 0.99, 0.999.\\
Fig.2 Energy spectrum of the fluctuation around the classical solution is 
plotted for $m$=1 and $k^2$=0.5.\\
Fig.3 The band structure of the spinless Fermion (the charge degree of 
freedom for finite $\delta$.\\
Fig.4 The region of ($q,\omega$) with a finite ${\rm Im}N(q,\omega)$ for several
choices of (a) $\delta=0.7$, (b) $\delta=0.3$, (c) $\delta=0.01$, (d) $\delta=0.0$ 
per one site.\\
Fig.5  The spectral density of the charge excitations, ${\rm Im}N(q,\omega)$, 
in the plane  of $q$ and $\omega$ for (a) $\delta=0.3$, (b) $\delta=0.1$, 
(c) $\delta=0.0$ per one site.\\
Fig.6 The Feynman diagram for the self-energy. The thick line means the full Green
function and the broken line represent the impurity scattering.\\
Fig.7 The density of states is plotted for $\zeta$=0, 0.01, 0.1, 0.3, 0.5, 1.2.\\
Fig.8 The compressibility is plotted for $\zeta$=0,0.01,0.1,0.3,0.5.\\
Fig.9 The spectral density of the charge excitations, ${\rm Im}N(q,\omega)$, 
in the disordered system as the function of $\omega$ for  (a) $\delta=0.0, 0.1,0.3,0.5, 
q=2$ and (b) $q=1, 2, 3, \delta=0.3$.\\
Fig.10 The optical-conductivity is plotted for $\delta$=0.01, and $\zeta$=0.0,
0.01,0.1,0.3,0.5.\\
Fig.11 The optical-conductivity is plotted for $\zeta$=0.1, and $\delta$=0.0,
0.01,0.1,0.3. 

\begin{thebibliography}{99}
\bibitem{hubbard} J. Hubbard: {Proc. Roy. London, {\bf A281} (1964) 401}. 
\bibitem{br} W. F. Brinkman and T. M. Rice: {Phys. Rev. {\bf B2} (1970) 4302}.
\bibitem{liebwu} E. H. Lieb and F. Y. Wu: {Phys. Rev. Lett. {\bf 20} (1968) 1445}.
\bibitem{usuki} T. Usuki, N. Kawakami and A. Okiji: {Phys. Lett. {\bf 135A} (1989) 476}.
\bibitem{shastry} B. S.Shastry and B. Sutherland: {Phys. Rev. Lett. {\bf 65} (1990) 243}.
\bibitem{kawakami} N. Kawakami and S. K. Yang: Phys. Lett. {\bf 148A}  (1990) 359.
\bibitem{korepin} H. Frahm and V. E. Korepin: Phys. Rev. {\bf B42} (1990) 10553. 
\bibitem{schulz} H. Schulz: Phys. Rev. Lett. {\bf 64} (1990) 2831.
\bibitem{tomonaga} S. Tomonaga: { Prog. Theor. Phys. {\bf 5} (1950) 349}. 
\bibitem{luttinger} J. M. Luttinger: { Phys. Rev. {\bf 4} (1963) 1154}.
\bibitem{emery} V. J. Emery: {\it Highly Conducting 
One-Dimensional Solids, eds. J. Devreese, R. Evrad and V. van Doren 
{\bf (}Plenum, New York, 1979{\bf )} p.247}.
\bibitem{solyom} J. Solyom: { Adv. Phys. {\bf 28} (1979) 201}.
\bibitem{fukuyama} H. Fukuyama and H.Takayama: {\it Electronic 
Properties of Inorganic Quasi-One-Dimensional Compounds Part 1, ed.  
P. Monceau {\bf (}D. Reidel Pub .Co., 1985{\bf )} p,41}.
\bibitem{emery1} V. J. Emery, A. Luther and I. Peschel: {Phys. Rev. B {\bf 13} (1976) 1272}.
\bibitem{emery2} V. J. Emery: {Phys. Rev. Lett {\bf 65} (1990) 1076}.
\bibitem{giamarchi} T. Giamarchi: {Phys. Rev. B {\bf 44} (1991) 2905}.
\bibitem{giamarchi2} T. Giamarchi: {Phys. Rev. B {\bf 46} (1992) 342}.
\bibitem{giamarchi3} T. Giamarchi and A. J. Millis: {Phys. Rev. B {\bf 46} (1991) 9325}.
\bibitem{emery3} V. J. Emery: {\it CORRELATED ELECTRON SYSTEMS Volume 9, 
ed. V. J. Emery {\bf (}World Scientific Pub. Co. Pte. Ltd, 1993{\bf )} p.166}.
\bibitem{mori} M.Mori, H.Fukuyama and M.Imada: {J. Phys. Soc. Jpn {\bf 63}(1994)1639}.
\bibitem{suzumura} Y. Suzumura: { Prog. Theor. Phys.  {\bf 61} (1979) 1}.
\bibitem{haldane} F. D. M. Haldane: { Phys. {\bf C 14} (1981) 2585}; 
 {Phys. Rev. Lett. {\bf 47} (1981) 1840}.
\bibitem{penc} K. Penc and J. Solyom: { Phys. Rev. {\bf B471} (1993) 6273}.
\bibitem{mcmillan} W. L. McMillan: Phys. Rev. {\bf B14} (1976) 1496; {\bf B16} (1977) 4655.
\bibitem{coleman} S. Coleman: { Phys. Rev. {\bf D11} (1975) 2088}.
\bibitem{mandelstam} S. Mandelstam: { Phys. Rev. {\bf D11} (1975) 3026}.
\bibitem{okwamoto} Y. Okwamoto: { J. Phys. Soc. Jpn {\bf 49} (1980) 8}. 
\bibitem{tm} S. Takada and S. Misawa: { Prog. Theor. Phys. {\bf 66} (1981) 101}. 
\bibitem{hida} K. Hida, M. Imada and M. Ishikawa: J. Phys. C: Solid State Phys. {\bf 16} (1983) 4945.
\bibitem{luther} A. Luther and V. J. Emery: { Phys. Rev. Lett. {\bf 33} (1974) 589}.
\bibitem{lee} P. A. Lee: {Phys. Rev. Lett. {\bf 34} (1975) 1247}.
\bibitem{sutherland} B. Sutherland: { Phys. Rev. {\bf A8} (1973) 2514}.
\bibitem{preuss} R. Preuss, A. Muramatsu, W. von der Lindon, F. F. Assaad and W. Hanke:
{Phys. Rev. Lett. {\bf 73} (1994) 732}.
\bibitem{imada}M. Imada, N. Furukawa and T. M. Rice: J. Phys. Soc. Jpn. {\bf 61} (1992) 3861.
\bibitem{haldane3}F. D. M. Haldane: {Phys. Rev. Lett. {\bf 74} (1995) 2090}.
\bibitem{suzumura-cardy} Y. Suzumura, T. Saso, H. Fukuyama and J. L. Cardy: 
{\it Proc. Int. Conf. LT-17, Karlsruche, 1984} 
 (North Holland Publishing Co., 1984), 891.
\bibitem{suzumurafukuyama}  Y. Suzumura and H. Fukuyama: J. Phys. Soc. Jpn. 
{\bf 53} (1984) 3918.
\bibitem{fl} H. Fukuyama and P. A. Lee: {Phys. Rev. {\bf B 17} (1978) 535}.
\end{thebibliography}
\end{document}